\documentclass[a4paper,11pt]{article}
\pdfoutput=1 

\usepackage{jcappub} 

\usepackage[T1]{fontenc} 

\title{Halo modelling in chameleon theories}

\author[a,b]{Lucas~Lombriser,}
\author[a]{Kazuya~Koyama,}
\author[c]{Baojiu~Li}

\affiliation[a]{Institute of Cosmology and Gravitation, University of Portsmouth, Dennis Sciama Building, Burnaby Road, Portsmouth, PO1~3FX, U.K.}
\affiliation[b]{Institute for Astronomy, University of Edinburgh, Royal Observatory, Blackford Hill, Edinburgh, EH9~3HJ, U.K.}
\affiliation[c]{Institute for Computational Cosmology, Ogden Centre for Fundamental Physics, Department of Physics, University of Durham, Science Laboratories, South Road, Durham, DH1 3LE, U.K.}

\emailAdd{lucas.lombriser@port.ac.uk}
\emailAdd{kazuya.koyama@port.ac.uk}
\emailAdd{baojiu.li@durham.ac.uk}

\newcommand{\tsec}[1]{Sec.~\ref{#1}}
\newcommand{\tsecs}[1]{Secs.~\ref{#1}}
\newcommand{\bq}{\begin{equation}}
\newcommand{\eq}{\end{equation}}
\newcommand{\bqa}{\begin{eqnarray}}
\newcommand{\eqa}{\end{eqnarray}}

\newcommand{\rmd}{\ensuremath{\mathrm{d}}}
\newcommand{\Msun}{M_\odot}
\newcommand{\Msunh}{M_\odot /h}

\newcommand{\Mpch}{h\,\text{Mpc}^{-1}}
\newcommand{\hGpc}{h^{-1}~\text{Gpc}}

\newcommand{\Sm}{S_{\rm m}}
\newcommand{\rhom}{\rho_{\rm m}}
\newcommand{\rhomb}{\bar{\rho}_{\rm m}}
\newcommand{\drhom}{\delta\rhom}
\newcommand{\Om}{\Omega_{\rm m}}
\newcommand{\Olam}{\Omega_{\Lambda}}
\newcommand{\scal}{\varphi}
\newcommand{\varscal}{\phi}
\newcommand{\bscal}{\bar{\scal}}
\newcommand{\dscal}{\delta\varphi}
\newcommand{\dscallin}{\dscal_{\rm lin}}

\newcommand{\Mvir}{M_{\rm vir}}
\newcommand{\rvir}{r_{\rm vir}}
\newcommand{\cvir}{c_{\rm vir}}
\newcommand{\rhos}{\rho_{\rm s}}
\newcommand{\rs}{r_{\rm s}}
\newcommand{\RTH}{r_{\rm th}}
\newcommand{\rhoin}{\rho_{\rm in}}
\newcommand{\rhoout}{\rho_{\rm out}}
\newcommand{\rcham}{r_{\rm cham}}
\newcommand{\yhal}{y_{\rm h}}
\newcommand{\yenv}{y_{\rm env}}
\newcommand{\deltac}{\delta_{\rm c}}

\newcommand{\denv}{\delta_{\rm env}}

\abstract{
We analyse modelling techniques for the large-scale structure formed in scalar-tensor theories of constant Brans-Dicke parameter which match the concordance model background expansion history and produce a chameleon suppression of the gravitational modification in high-density regions.
Thereby, we use a mass and environment dependent chameleon spherical collapse model,
the Sheth-Tormen halo mass function and linear halo bias, the Navarro-Frenk-White halo density profile, and the halo model.
Furthermore, using the spherical collapse model, we extrapolate a chameleon mass-concentration scaling relation from a $\Lambda$CDM prescription calibrated to $N$-body simulations.
We also provide constraints on the model parameters to ensure viability on local scales.
We test our description of the halo mass function and nonlinear matter power spectrum
against the respective observables extracted from large-volume and high-resolution $N$-body simulations in the limiting case of $f(R)$ gravity, corresponding to a vanishing Brans-Dicke parameter.
We find good agreement between the two; the halo model provides a good qualitative description of the shape of the relative enhancement of the $f(R)$ matter power spectrum with respect to $\Lambda$CDM caused by the extra attractive gravitational force but fails to recover the correct amplitude.
Introducing an effective linear power spectrum in the computation of the two-halo term to account for an underestimation of the chameleon suppression at intermediate scales in our approach, we accurately reproduce the measurements from the $N$-body simulations.
}

\begin{document}
\maketitle
\flushbottom


\section{Introduction} \label{sec:intro}

In the effective field theory limit at low energies, models attempting to unify general relativity with the standard model interactions are typically expected to introduce a scalar field in addition to the gravitational tensor field, which may couple minimally or nonminimally to the matter fields.
This fifth element can source the observed late-time acceleration of our Universe as an alternative to the cosmological constant (cf.~\cite{wang:12}).
A nonminimal coupling consequently leads to a modification of the gravitational interactions between the matter fields, which is, however, tightly constrained by local observations~\cite{will:05}.
If the scalar field potential has an adequate form, it can contribute to alleviate these constraints such as is the case in chameleon models~\cite{khoury:03a,khoury:03b,brax:04,cembranos:05},
where the curvature dependence of the scalar field is such to suppress the extra force in high-density regions.
At low curvature and below the Compton wavelength of the scalar field, the gravitational force remains enhanced, yielding an increase in the growth of structure.

Here, we specialise to scalar-tensor models with constant Brans-Dicke parameter $\omega$ that match the $\Lambda$CDM background expansion history, exhibit a chameleon mechanism, and reduce to the Hu-Sawicki~\cite{hu:07a} $f(R)$ model~\cite{buchdahl:70,starobinsky:79,starobinsky:80,capozziello:03,carroll:03,nojiri:03} in the limit of $\omega=0$.
The enhanced gravitational coupling of these models at low curvature and below the Compton wavelength can be utilised to place observational constraints on the gravitational modifications.
Such constraints from current and expected from future observations have been particularly well studied in the limiting case of $f(R)$ gravity~\cite{zhang:05,song:06,li:07,zhang:07,song:07,hu:07a,jain:07,brax:08,zhao:08,schmidt:09,giannantonio:09,reyes:10,lombriser:10,yamamoto:10,wang:10,ferraro:10,jain:11,hojjati:11,gilmarin:11,wojtak:11,lombriser:11b,terukina:12,divalentino:12,jain:12,samushia:12,he:12,hu:12,okada:12,hall:12,lombriser:13a,brax:13,marchini:13a,abebe:13,lam:13,hellwing:13,upadhye:13,marchini:13b,hu:13,sakstein:13,zu:13,cai:13,baldi:13,arnold:13}.
Currently, the strongest bounds on these modifications are inferred from the transition required to interpolate between the low curvature of the large-scale structure and the high curvature of the galactic halo~\cite{hu:07a} as well as from the comparison of nearby distance measurements in a sample of unscreened dwarf galaxies~\cite{jain:12}.
Hereby, in order to guarantee chameleon screening, the background field amplitude has to be smaller than the corresponding depth of the potential wells, $|\Psi|\sim(10^{-7}-10^{-5})$.
Independently, strong observational constraints have also been inferred from the large-scale structure such as from the analysis of cluster profiles~\cite{lombriser:12} and abundance~\cite{schmidt:09,lombriser:10,ferraro:10}, which are however, still 2-3 orders of magnitude weaker than the bounds inferred from the local and astrophysical tests.
Note, however, that while astrophysical and Solar System tests require the chameleon field to couple to the baryonic components, due to the dominance of the dark matter on cosmological scales, probes of the large-scale structure are typically independent of this assumption.
Such dark chameleon fields that only couple to dark matter may furthermore evade problems arising in the early universe due to high-energy fluctuations through quantum particle production~\cite{erickcek:13a,erickcek:13b}.
The current cosmological bounds have been obtained from an analysis of massive clusters, implying constraints that lie in a regime where the modified observables can correctly be described by performing a linearisation of the scalar field potential.
This approach breaks down for clusters with gravitational potential wells of the order of $|\Psi|\lesssim10^{-5}$, for which the chameleon mechanism becomes important and needs to be incorporated correctly.

While $N$-body simulations of such scalar-tensor theories provide a great laboratory for studying the chameleon mechanism~\cite{oyaizu:08a,li:09,li:10,li:11,puchwein:13,llinares:13b}, they are computationally expensive and more efficient modelling of the large-scale structure needs to be developed based on these simulations in order to allow for a full exploration of the cosmological parameter space in the model comparison to observations.
In this paper, we provide simple modelling techniques for the large-scale structure produced in chameleon theories, providing an important tool for efficiently extrapolating and interpolating the nonlinear quantities extracted from $N$-body simulations beyond the simulated values of the cosmological and chameleon model parameters implemented.
These tools are essential to allow the consistent inference of model constraints from the observed large-scale structure, enabling sufficient and smooth variation of chain parameters as well as statistical convergence. 


In \tsec{sec:theory}, we give a short review on scalar-tensor theories with a particular focus on chameleon models.
We generalise the Hu-Sawicki $f(R)$ chameleon model to scalar-tensor theories with constant Brans-Dicke parameter and discuss the background and Solar System constraints on these models.
In \tsec{sec:structure}, we describe the formation of large-scale structure in chameleon models using linear cosmological perturbation theory in the quasistatic limit, the mass and environment dependent spherical collapse model~\cite{li:11a}, methods for environmental averaging, and dark matter $N$-body simulations.
In \tsec{sec:halomodel}, we describe the halo model for chameleon theories and analyse its performance against $N$-body simulations in the limiting case of $f(R)$ gravity.
In specific, we follow the computation of Ref.~\cite{lombriser:13b} and use the Sheth-Tormen prescription~\cite{sheth:99} to obtain the halo mass function and extend this approach to the description of the linear halo bias.
Based on the chameleon spherical collapse model, we introduce a chameleon transition in the scaling of concentration with respect to mass and use this to describe the halo density profile assuming a Navarro-Frenk-White (NFW)~\cite{navarro:95} profile, which has been shown to provide good fits to chameleon halos in the limit of $f(R)$ gravity~\cite{lombriser:12}.
Combining these quantities in the halo model, we compute the nonlinear matter power spectrum for chameleon theories.
We test our description against dark matter $N$-body simulations of chameleon $f(R)$ gravity models, finding good agreement between the two;
the halo model provides a good qualitative description of the chameleon suppression in the nonlinear matter power spectrum but fails to reproduce the correct amplitude of the relative enhancement due to the chameleon field.
By introducing a transition between the linear $f(R)$ and $\Lambda$CDM power spectrum in the two-halo term to account for an underestimation of the chameleon suppression on intermediate scales, we improve the halo model description to obtain an accurate fit to the nonlinear matter power spectrum extracted from $N$-body simulations over a wide range of scales.
Finally, in \tsec{sec:conclusion}, we conclude with a discussion of our results.


\section{Chameleon model} \label{sec:theory}

We consider scalar-tensor theories for which the modified Einstein-Hilbert action in the Jordan frame can be written in the form
\bq
 S = \frac{1}{2\kappa^2} \int d^4x \sqrt{-g} \left[ F(\scal) R - Z(\scal) \partial^{\mu} \scal \, \partial_{\mu} \scal - 2 U(\scal) \right] + \Sm[\psi_{\rm m}; g_{\mu\nu}], \label{eq:jordanaction}
\eq
where $\kappa^2 \equiv 8 \pi \, G$ with the bare gravitational coupling $G$, $R$ is the Ricci scalar, $\Sm$ is the matter action with matter fields $\psi_{\rm m}$, and we have set the speed of light in vacuum to unity.
The scalar field $\scal$ is coupled to the metric $g_{\mu\nu}$ via $F(\scal)$ with kinetic coupling $Z(\scal)$ and scalar field potential $U(\scal)$.
It can be redefined to reduce the number of free functions of $\scal$ in~Eq.~(\ref{eq:jordanaction}) to two instead of three.
In this paper, we will use the Brans-Dicke~\cite{brans:61} representation for scalar-tensor models,
\bq
 F\equiv\scal, \hspace{8mm} Z\equiv\frac{\omega(\scal)}{\scal}. \label{eq:bransdicke}
\eq
We can recast the Jordan frame action Eq.~(\ref{eq:jordanaction}) into the Einstein frame by a conformal transformation of the metric such that
\bq
 S = \int d^4x \sqrt{-\tilde{g}} \left[ \frac{\tilde{R}}{2\kappa^2} - \frac{1}{2}\tilde{\partial}^{\mu}\varscal\,\tilde{\partial}_{\mu}\varscal - V(\varscal) \right] + \Sm[\psi_{\rm m}; A^2(\varscal) \tilde{g}_{\mu\nu}], \label{eq:einsteinaction}
\eq
where here and throughout the paper, tildes denote quantities in the Einstein frame, and
\bqa
 \tilde{g}_{\mu\nu} & \equiv & \scal\,g_{\mu\nu}, \\
 \left(\frac{\rmd \varscal}{\rmd \scal}\right)^2 & \equiv & \frac{1}{2\kappa^2} \frac{3+2\omega}{\scal^2}, \label{eq:scalrel} \\
 A(\varscal) & \equiv & \scal^{-1/2}, \\
 V(\varscal) & \equiv & \frac{U(\scal)}{\kappa^2\scal^2}.
\eqa
Integration of Eq.~(\ref{eq:scalrel}) relates the Einstein frame to the Jordan frame scalar field,
\bq
 \varscal = \frac{1}{\kappa}\sqrt{\frac{3+2\omega}{2}} \ln\scal + \varscal_0,
\eq
where we restricted to cases with Brans-Dicke parameter $\omega=\textrm{const.}$ and set $\varscal_0\equiv0$.
This shall also be adopted in the following discussion along with the condition that $\omega>-3/2$ to evade ghost fields.
Variation of the action Eq.~(\ref{eq:einsteinaction}) with respect to $\varscal$ yields the scalar field equation
\bq
 \tilde{\Box}\varscal = \frac{\kappa}{\sqrt{6+4\omega}} \tilde{T} + V'(\varscal) \equiv V'_{\rm eff}(\varscal), \label{eq:sfeq}
\eq
where $V_{\rm eff}(\varscal)$ is an effective potential governing the dynamics of $\varscal$ and the energy-momentum tensor is given by $\tilde{T}=A(\varscal)^4 T$.
For a scalar field with $\scal\simeq1$ minimising the effective potential, $V_{\rm eff}'(\varscal)=0$, Eq.~(\ref{eq:sfeq}) becomes
\bq
 \frac{\rmd}{\rmd\scal} U(\scal) \simeq \frac{1}{2}(\kappa^2\rhom+4U)\simeq\frac{\tilde{R}}{2}\simeq\frac{R}{2}, \label{eq:mincond}
\eq
where we have assumed dominance of the matter energy density $\rhom$ and further required $(\kappa^2\rhom+4U)\gg(3+2\omega)(\partial_{\mu}\scal)^2/2$.


\subsection{Background expansion history} \label{sec:background}

We are interested in scalar-tensor theories that predict a background expansion history that closely matches the one of $\Lambda$CDM.
Combining the Jordan frame scalar field equation and Einstein equations in the background, we obtain the Friedmann equations
\bqa
 3\bar{\scal}\,H^2 - \kappa^2\rhomb - \frac{H^2}{2}\frac{\omega}{\bar{\scal}}\bar{\scal}'^2 + 3H^2\bar{\scal}' - \bar{U} & = & 0, \label{eq:friedmann1} \\
 \omega H^2 \bar{\scal}'' + H^2 \left[ \omega\left(\frac{H'}{H}+3\right) - 3 \right]\bar{\scal}' - 3H^2\left[ \frac{H'}{H}+3 \right]\bar{\scal} + \kappa^2\rhomb + \bar{U} + \bscal\frac{\rmd \bar{U}}{\rmd \bar{\scal}} & = & 0, \label{eq:friedmann2}
\eqa
where here and throughout the paper, primes denote derivatives with respect to $\ln a$ and overbars refer to quantities evaluated at their background.
For $\bscal\simeq1$ and $|\omega^j\bscal^{(i)}|\ll1$ with order of the derivative $i=1,2$ and exponent $j=0,1$,
we can neglect these derivative terms of the scalar field in Eqs.~(\ref{eq:friedmann1}) and (\ref{eq:friedmann2}),
and it then follows directly that $\bar{U}_{\bscal} \simeq \bar{R}/2$.
For $|\bscal^{(i)}|\sim|\bscal-1|$, $|\omega|\ll|\bscal_0-1|^{-1}$, and $\bar{U}\simeq\Lambda$, $H^2$ is described by a $\Lambda$CDM expansion history at order $\mathcal{O}(|\bar{\scal}-1|)$.


\subsection{Chameleon regime} \label{sec:chameleon}

In high-density regions, where $\kappa\,\rhom \gg -\sqrt{6+4\omega}\tilde{\nabla}^2\varscal$ in the quasistatic limit, Eq.~(\ref{eq:sfeq}) implies that the scalar field potential satisfies $U_{\scal} \simeq R/2$ as in Eq.~(\ref{eq:mincond}).
We make the ansatz $U=\Lambda + U_{\alpha} (1-\scal)^{\alpha}$ with positive constant $\alpha$ such that we recover $\Lambda$CDM in the limit of $\scal=1$.
With the requirement that $U_{\scal} \simeq R/2$, hence $\alpha\neq1$, we obtain
\bqa
 U & \simeq & \Lambda -\frac{\bar{R}_0}{2\alpha} \frac{(1-\scal)^{\alpha}}{(1-\bscal_0)^{\alpha-1}} \simeq \Lambda + \frac{\bscal_0-1}{2\alpha}\left(\frac{\bar{R}_0}{R^{\alpha}}\right)^{1/(1-\alpha)}, \label{eq:scalpot} \\
 \scal & \simeq & 1 + (\bscal_0-1) \left(\frac{\bar{R}_0}{R}\right)^{1/(1-\alpha)}, \label{eq:scalfield}
\eqa
where here and throughout the paper, subscripts of zero refer to present time, $a\equiv1$.
Since in high-density regions, $R \gg \bar{R}_0$, for $\alpha<1$, we get $(\scal-1) \simeq 2\kappa/\sqrt{6+4\omega}\,\varscal\simeq0$.
Consequently, modifications of gravity are suppressed.
We further require $\alpha\gg|\bscal_0-1|$ such that $\bar{U}\simeq\Lambda$ at the background and $|\omega^j||1-\bscal_0|\ll(1-\alpha)^2$ such that $|\omega^j\scal^{(i)}|\ll1$.
This satisfies the conditions for $\scal$ imposed in \tsec{sec:background}.

Note that we can alternatively write the scalar field Eq.~(\ref{eq:scalfield}) and its potential Eq.~(\ref{eq:scalpot}) replacing
$R$ with $-\kappa^2 T_{\rm eff}\simeq(-\kappa^2T+4\Lambda)\simeq(\kappa^2\rhom+4\Lambda)$ or $3(H^2+\Lambda)$ at the background,
where $T_{\rm eff}$ is an effective energy-momentum tensor including the dark energy contributions.
Furthermore, in the limit of $\omega\equiv0$, defining $\scal\equiv1+\rmd f/\rmd R \equiv 1+f_R$, the action Eq.~(\ref{eq:jordanaction}) in the parametrisation Eq.~(\ref{eq:bransdicke}) and with scalar field potential Eq.~(\ref{eq:scalpot}) reduces to the Hu-Sawicki $f(R)$ gravity model~\cite{hu:07a} with $\alpha \equiv n/(n+1)$.
Hence, Eq.~(\ref{eq:scalpot}) shall serve here as a straightforward generalisation thereof with model parameters $|\bscal_0-1|$, $\omega$, and $\alpha$.


\subsection{Linearised regime} \label{sec:linear}

In the following, we focus on the quasistatic limit, require $\scal\simeq1$, and linearise the contribution of $\scal$ to the scalar field potential in Eq.~(\ref{eq:sfeq}) with respect to the background.
Subsequently subtracting the background, we obtain
\bq
 \tilde{\nabla}^2 \dscallin - m^2 \dscallin + \frac{\kappa^2}{3+2\omega} \delta\rhom \simeq 0, \label{eq:qssfeq}
\eq
where we have defined
\bq
 m^2 \equiv \frac{1-\alpha}{3+2\omega}\frac{(1-\bar{\scal})^{\alpha-2}}{(1-\bar{\scal}_0)^{\alpha-1}}\bar{R}_0 \simeq \frac{1-\alpha}{3+2\omega}\frac{(\bar{R}_0\bar{R}^{\alpha-2})^{1/(\alpha-1)}}{1-\bscal_0},
 \label{eq:scalmass}
\eq
as well as $\dscal \equiv \scal - \bscal$ and $\delta\rhom \equiv \rhom - \bar{\rho}_{\rm m}$.

In \tsec{sec:structure}, we will be interested in the large-scale structure formed in chameleon theories.
Dark matter halos extracted from $\Lambda$CDM $N$-body simulations are well described assuming sphericity of the halos and that their density profiles are given by the NFW~\cite{navarro:95} expression
\bq
 \delta\rhom(r) = \frac{\rho_{\rm s}}{\frac{r}{r_{\rm s}} \left( 1+\frac{r}{r_{\rm s}} \right)^2 }, \label{eq:nfw}
\eq
where $\rhos$ and $\rs$ are the characteristic density and scale, respectively, which are fitted to the simulations.
It has been shown in Ref.~\cite{lombriser:12} that the NFW fit also correctly describes the dark matter halo density profiles produced in $N$-body simulations of chameleon $f(R)$ gravity.
Motivated by these results, we assume here that it also provides a good description for halo density profiles formed in more general chameleon models defined by the action Eq.~(\ref{eq:jordanaction}) and the scalar field potential Eq.~(\ref{eq:scalpot}).
Following the derivation of Ref.~\cite{lombriser:12} under this assumption, the linearised scalar field in the virialised halo is then determined by
\bqa
 \dscallin & \simeq & -\frac{\kappa^2\rhos\rs^3}{6+4\omega} \left\{ \Gamma[0,m(r+\rs)] e^{2m(r+\rs)} + \Gamma[0,-m(r+\rs)] \right. \nonumber\\
 & & \left. - \Gamma(0,-m\,\rs) - e^{2m\,\rs}\Gamma(0,m\,\rs) \right\} \frac{e^{-m(r+\rs)}}{r},
\eqa
where
\bq
 \Gamma(s,r) = \int_r^{\infty} \rmd t \, t^{s-1} e^{-t}
\eq
is the upper incomplete gamma function.
In the limit of $\rhom\gg\rhomb$, the linearised scalar field becomes
\bq
 \dscallin \simeq \frac{\kappa^2 \rho_{\rm s} r_{\rm s}^3}{3+2\omega} \left[ \frac{\ln (1+r/r_{\rm s})}{r} - m\,e^{m\,r_{\rm s}} \Gamma(0,m\,r_{\rm s}) \right]. \label{eq:hdlinsf}
\eq


\subsection{Intermediate regime} \label{sec:intermediate}

In \tsec{sec:chameleon}, we have described the chameleon screening of the scalar field in high-density regions and in \tsec{sec:linear}, the linearised limit of the scalar field when deviations from the background are small.
Next, we are interested in the description of the radial profile of the scalar field $\scal$ in the intermediate regime, interpolating between the linearised and screened chameleon limits.

We begin by studying the case of a constant spherically symmetric top-hat matter density $\rhoin$ of radius $\RTH$, embedded in an outer matter density $\rhoout$.
On the inside and outside of $\RTH$, the effective Einstein frame scalar field potential $V_{\rm eff}(\varscal)$ in Eq.~(\ref{eq:sfeq}) is minimised and we refer to the corresponding Jordan frame scalar field as $\scal_{\rm in}$ and $\scal_{\rm out}$, respectively.
The distance $\Delta r\geq0$ that is necessary for $\scal\simeq1$ to settle from $\scal_{\rm out}$ to $\scal_{\rm in}$ can be approximated by~\cite{khoury:03b}
\bq
 \frac{\Delta r}{\RTH}
 \simeq (3+2\omega) \frac{\scal_{\rm in} - \scal_{\rm out}}{6\Psi_{\rm N}}
 = (3+2\omega) \frac{\scal_{\rm in} - \scal_{\rm out}}{\kappa^2 \rho_{\rm in}\RTH^2},
\eq
where we have used the Newtonian potential at the surface of the sphere,
\bq
 \Psi_{\rm N} = \frac{\kappa^2}{8\pi} \frac{M}{\RTH} = \frac{\kappa^2}{6} \rhoin \RTH^2
\eq
with mass $M \equiv 4\pi \, \rhoin \RTH^3$.
The inner and outer solutions for $\scal$ are obtained from $U_{\scal} \simeq R/2$ as described in Eq.~(\ref{eq:scalfield}) for the chameleon models of interest here.
Hence, we have
\bq
 \scal_{\rm in/out} \simeq 1 + (\bscal_0-1)\left[ \frac{1+4 \frac{\Olam}{\Om}}{\tilde{\rho}_{\rm in/out} a^{-3} + 4 \frac{\Olam}{\Om}} \right]^{1/(1-\alpha)},
\eq
where $\tilde{\rho}_{\rm in/out} \equiv \rho_{\rm m, in/out}(a=1)/\bar{\rho}_{\rm m}(a=1)$,
such that
\bq
 \frac{\Delta r}{\RTH} \simeq \frac{(3+2\omega)(\bscal_0-1) a^3}{3\Om \tilde{\rho}_{\rm in} (H_0 \RTH)^2} \left[ \left( \frac{1 + 4\frac{\Olam}{\Om}}{\tilde{\rho}_{\rm in} a^{-3} + 4\frac{\Olam}{\Om}} \right)^{\frac{1}{1-\alpha}} - \left( \frac{1 + 4\frac{\Olam}{\Om}}{\tilde{\rho}_{\rm out} a^{-3} + 4\frac{\Olam}{\Om}} \right)^{\frac{1}{1-\alpha}} \right]. \label{eq:thsh}
\eq
For a thin shell, $\Delta r=\RTH-r_0\ll \RTH$, the intermediate scalar field within $r\in[r_0,\RTH]$ is described via~\cite{khoury:03b, li:11a, lombriser:13b}
\bq
 \scal(r) \simeq \scal_{\rm in} - \frac{1}{3+2\omega} \frac{\kappa^2\rho_{\rm in}}{3} \left( \frac{r^2}{2} + \frac{r_0^3}{r} - \frac{3}{2} r_0^2 \right),
\eq
for which the force enhancement $\Delta F \equiv F-F_{\rm N}$ at $\RTH$ with total force $F$, Newtonian force $F_{\rm N} = G\,m\,M/\RTH^2$, and test mass $m_{\rm t}$ due to the extra coupling becomes
\bqa
 \frac{\Delta F}{F_{\rm N}} & \equiv & \left. -\frac{m_{\rm t}}{2F_{\rm N}} \tilde{\nabla} \scal \right|_{\RTH}
 \simeq \frac{1}{3+2\omega} \left[ 1 - \left( \frac{r_0}{\RTH} \right)^3 \right] \nonumber\\
 & = & \frac{1}{3+2\omega} \left[ 3\frac{\Delta r}{\RTH} - 3\left(\frac{\Delta r}{\RTH}\right)^2 + \left(\frac{\Delta r}{\RTH}\right)^3 \right]. \label{eq:extraforce}
\eqa
As for $r_0\ll\RTH$, we reproduce the force enhancement in the thick-shell regime $\Delta r>\RTH$~\cite{lombriser:13b,li:11a}, we apply Eq.~(\ref{eq:extraforce}) to all $\Delta r$.
Hence, we get
\bq
 \frac{\Delta F}{F_{\rm N}} \simeq \frac{1}{3+2\omega}\min\left[ 3\frac{\Delta r}{\RTH} - 3\left(\frac{\Delta r}{\RTH}\right)^2 + \left(\frac{\Delta r}{\RTH}\right)^3, 1 \right], \label{eq:enhf}
\eq
yielding an interpolation between the suppressed regime $\Delta F=0$ and the $\Delta F/F_{\rm N}=(3+2\omega)^{-1}$ enhancement, which is $\mathcal{C}^0$ for $\Delta r/\RTH\rightarrow0$ and $\mathcal{C}^2$ for $\Delta r/\RTH\rightarrow1$.

We will use the force enhancement Eq.~(\ref{eq:extraforce}) in the spherical collapse model when studying the formation of structure in chameleon theories assuming that dark matter halos are top hats.
We can, however, model the scalar field within dark matter halos more accurately by assuming a fit of the halo density profile as we have done in \tsec{sec:linear}, where we considered the linearised scalar field $\dscallin$ within a virialised cluster.
Assuming spherical symmetry in the quasistatic limit of the scalar field equation, Eq.~(\ref{eq:sfeq}), and that $\drhom$ is given by a NFW profile, one obtains a second-order differential equation, which can easily be integrated numerically using the substitution $\scal-1 = - e^{u(r)}$ (see, e.g.,~\cite{oyaizu:08a,zhao:10b,lombriser:12}).
Alternatively, the chameleon transition can be modelled following the semi-analytic approach of Pourhasan \emph{et al.}~\cite{pourhasan:11} for describing the chameleon field by matching the chameleon interior solution of \tsec{sec:chameleon}, applying to $r\in(r_-,r_+)$, to the chameleon exterior solution of \tsec{sec:linear} for $r > \rcham$ at the transition scale $\rcham$.
More precisely, the integration constants obtained from the integration of the quasistatic scalar field equation in the limit of $\rhom\gg\rhomb$, i.e., when the term $-m^2\dscal$ can be neglected in Eq.~(\ref{eq:qssfeq}), are defined by matching $\scal$ at the background and requiring that the matched scalar field and its derivative are continuous at the transition:
\bq
 \dscal(r) \equiv \left( \dscal_{r<\rcham}^{\rm in} \cup \dscal_{r \geq \rcham}^{\rm out} \right)(r) \in \mathcal{C}^1(U)
\eq
with $\rcham \in U \subset \mathbb{R}^+_0$.
The transition scale $\rcham$ is then computed numerically.
Finally, we can also assume an instantaneous transition to $\dscal=1-\bscal$ simply implemented via~\cite{lombriser:12}
\bq
 \scal \approx \min \left(\scal_{\rm lin}, 1 \right) \label{eq:instrans}
\eq
or equivalently, $\dscal \approx \min \left( \dscallin, 1-\bscal \right)$, using the linearised scalar field $\scal_{\rm lin}$ described in \tsec{sec:linear}.
All of these approaches have been shown to provide good agreement with $N$-body simulations of chameleon $f(R)$ gravity~\cite{lombriser:12}.
It is important to note, however, that matching $\dscal$ to boundary values such as $\dscal(\rvir)$, where $\rvir$ is the virial radius, given from simulations is essential for recovering the radial profile of the simulated scalar field.

Using the approximation of an instantaneous chameleon transition Eq.~(\ref{eq:instrans}) in the limit $\rhom\gg\rhomb$, where $\dscallin$ is described by Eq.~(\ref{eq:hdlinsf}), the scale of the chameleon transition $\rcham$ can be estimated by solving $\dscallin(\rcham)\approx1-\bar{\scal}$, which yields
\bq
 \rcham \simeq -r_{\rm s} - A^{-1} W\left[-A\,r_{\rm s} \exp(-A\,r_{\rm s})\right],
\eq
where $W[\cdot]$ is the Lambert $W$ function solving $x=W(x)\exp[W(x)]$ and we have defined
\bq
 A \equiv \frac{3+2\omega}{\kappa^2 \rho_{\rm s} r_{\rm s}^3}(1-\bscal) + m\, e^{m\, r_{\rm s}} \Gamma(0,m\,r_{\rm s}).
 \label{eq:lambertA}
\eq


\subsection{Solar System constraints} \label{sec:locconst}

\begin{figure}
 \centering
 \resizebox{\hsize}{!}{
  \resizebox{0.5015\hsize}{!}{\includegraphics{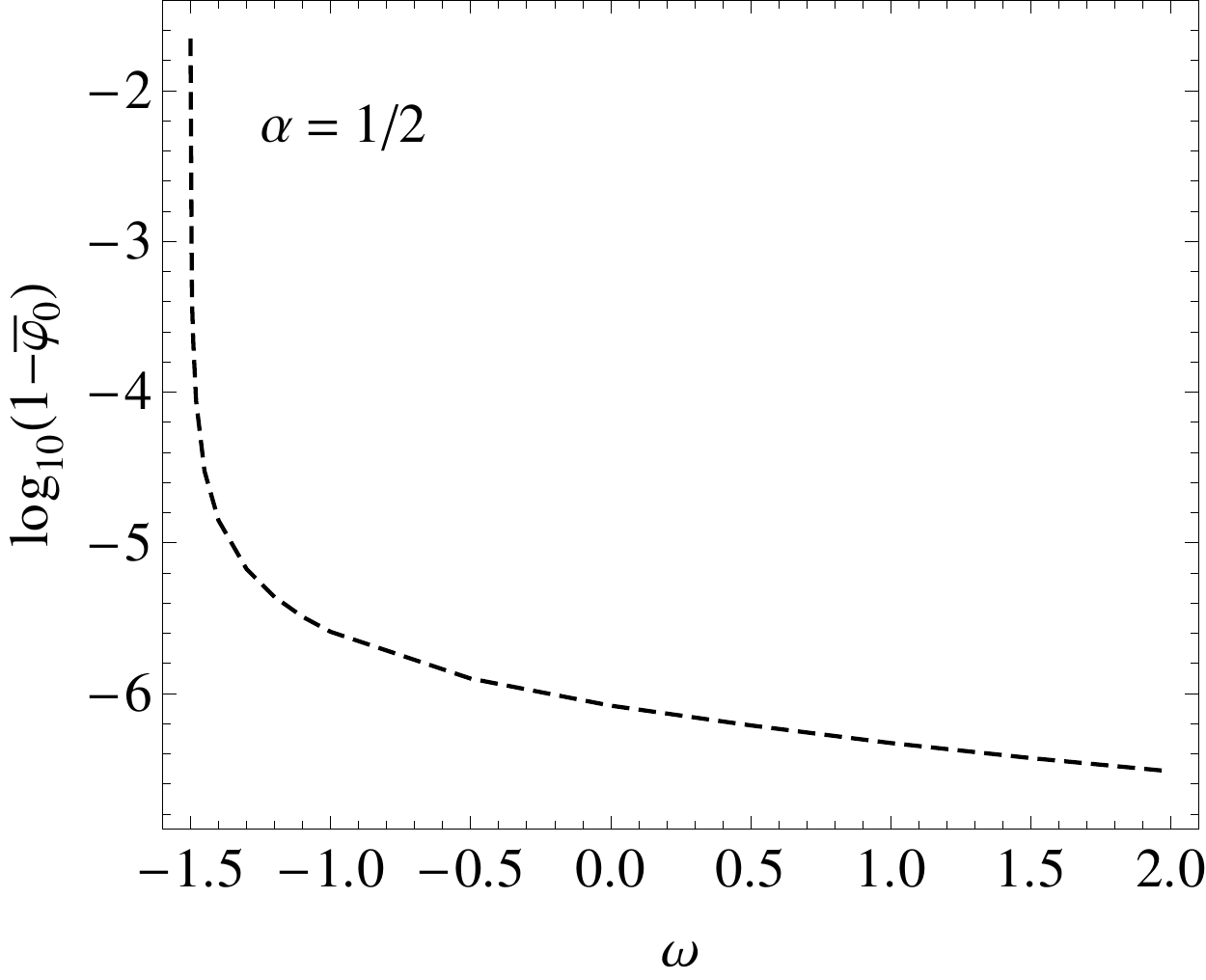}}
  \resizebox{0.4985\hsize}{!}{\includegraphics{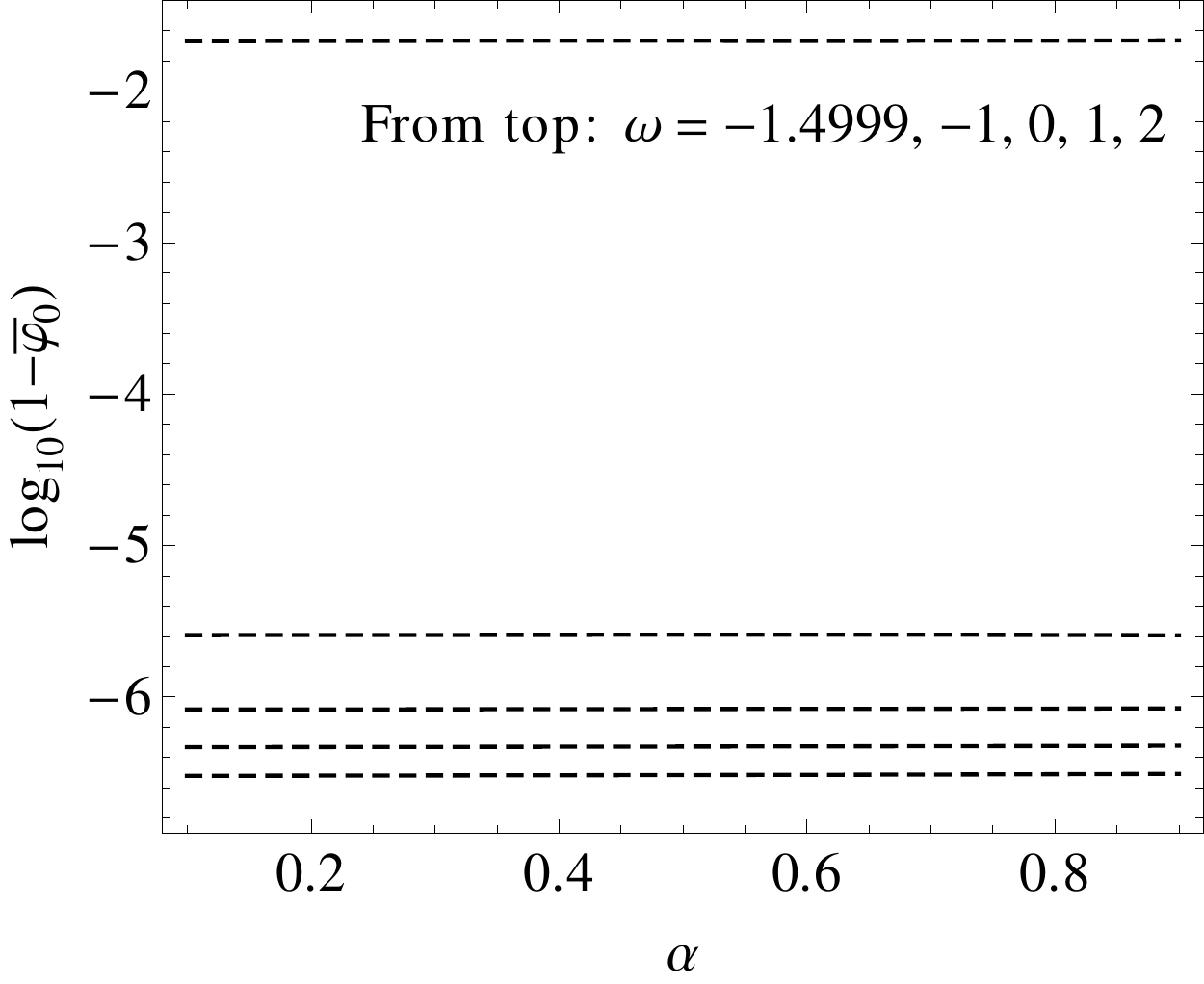}}
 }
 \caption{
  Upper bounds on $\log_{10}(1-\bscal_0)$ ensuring that the chameleon mechanism screens the Solar System from modifications of gravity.
  Hereby, $\bscal_0$ denotes the present cosmological background value of the scalar field.
  \emph{Left panel:} The constraint depends on the Brans-Dicke parameter $\omega$, which is assumed constant.
  \emph{Right panel:} Variation of $\alpha$, which enters the scalar field potential through $|U-\Lambda|\sim|\scal-1|^{\alpha}$, only affects the constraints very weakly through the scalar field mass $m$, Eq.~(\ref{eq:scalmass}), in Eq.~(\ref{eq:lambertA}).
  }
\label{fig:constraints}
\end{figure}

Before continuing with the discussion of the formation of structure within chameleon theories in \tsec{sec:structure}, we provide a short and simplified analysis of the requirements on our chameleon models given through Eqs.~(\ref{eq:jordanaction}) and (\ref{eq:scalpot}) to be viable within the Solar System.
Using the modelling of the scalar field within a dark matter halo described in \tsec{sec:intermediate}, we estimate Solar System constraints on the model parameter $|1-\bscal_0|$ as a function of $\omega$ and $\alpha$.
Hereby, in order to satisfy the tight local constraints on deviations from general relativity~\cite{will:05}, we simply require that the Milky Way halo with mass $M_{\rm MW}\approx1.26\times10^{12}~\Msun$~\cite{mcmillan:11} is screened within $\rcham\approx8~\textrm{kpc}$, where the Solar System is approximately located.

Note that we conservatively assume that the measured value of $M_{\rm MW}$ also applies to the case of modified gravity.
As in viable scalar-tensor theories, due to the enhanced gravitational force comparable to Eq.~(\ref{eq:enhf}), dynamically inferred masses are larger than lensing masses, which we use to determine the chameleon screening, the true halo mass can only be smaller than $M_{\rm MW}$, implying stronger constraints on $|\bscal_0-1|$ in order to achieve screening at the Solar System scale.
Hence, using the Milky Way mass inferred assuming general relativity weakens the constraint on $|\bscal_0-1|$.

For simplicity, we furthermore neglect the bulge and disc and assume that the NFW profile describes the dark matter profile of the Milky Way halo and galaxy sufficiently well for our approximations as well as that the environment can be approximated by the cosmological background.
Note that at the scale of the Solar System, the baryonic components dominate over the dark matter
and that the inclusion of the baryonic distribution may alleviate constraints on the chameleon field amplitude.
As we are only interested in an approximative bound, we choose to neglect the baryonic contribution and leave a more accurate derivation of the Solar System constraints for future work.
We model the NFW parameters in Eq.~(\ref{eq:nfw}) via the relations
\bqa
 \rhos & = & \frac{1}{3} \bar{\rho}_{\rm m} \Delta_{\rm vir} \cvir^3 \left[ \ln(1+\cvir) - \frac{\cvir}{\cvir+1} \right]^{-1}, \label{eq:chardens} \\
 \rs & = & \frac{1}{\cvir}\left( \frac{3\Mvir}{4\pi\bar{\rho}_{\rm m}\Delta_{\rm vir}} \right)^{1/3}, \label{eq:charscale}
\eqa
where $\cvir\equiv\rvir/\rs$ is the virial halo concentration, $\Delta_{\rm vir}$ is the virial overdensity, and $\Mvir$ is the virial halo mass.
We assume the cosmological parameter and overdensity values defined in \tsec{sec:sims} and follow Refs.~\cite{schmidt:08,lombriser:11b} to model the concentration, generalising this approach beyond the $\omega=0$ case of $f(R)$ gravity with the computation of the linear matter power spectrum as described in \tsec{sec:lingrow}.
We refer to \tsec{sec:concprof} for more details on the modelling of $\cvir$, $\rhos$, and $\rs$.

In Fig.~\ref{fig:constraints}, we show the Solar System constraints the approximation $\dscal_{\rm lin,0}(\rcham)\approx1-\bar{\scal}_0$ implies on $|\bscal_0-1|$ as a function of $\omega$ and $\alpha$.
While the constraints do only depend very weakly on the exponent of the scalar field potential $\alpha$, they scale with the Brans-Dicke parameter $\omega$ approximately as
\bq
 |\bscal_0-1| \lesssim \frac{5}{6+4\omega} \times 10^{-6}. \label{eq:solsyscon}
\eq
For $f(R)$ gravity, where $\omega=0$, this implies that $|\bar{f}_{\bar{R}0}|=|\bscal_0-1|\lesssim8\times10^{-7}$, which is in agreement with Ref.~\cite{hu:07a}.

Note that when neglecting the integration constant in Eq.~(\ref{eq:hdlinsf}), we can also write the constraint in Eq.~(\ref{eq:solsyscon}) as $|\scal_0-1|\lesssim 2|\Psi_{\rm N}|/(3+2\omega)$.
We can furthermore combine the constraint in Eq.~(\ref{eq:solsyscon}), assuming equality, with the background constraints assumed in \tsec{sec:background}.
The condition that $\bar{U}_0\approx\Lambda$ implies that
\bq
 \frac{1}{6+4\omega} \ll 10^5\alpha.
\eq
Hence, the modification satisfies $\alpha\gg|\bscal_0-1|$ assumed in \tsec{sec:chameleon} and $|\omega|\ll|\bscal_0-1|^{-1}$ assumed in \tsec{sec:background}.
Finally, note that, in general, the constraint in Eq.~(\ref{eq:solsyscon}) does not apply to chameleon fields which couple differently to the baryons than to the dark matter field.


\section{Structure formation in the presence of a chameleon field} \label{sec:structure}

In the following, we study the formation and evolution of structure in the cold dark matter scenarios of $\Lambda$CDM and scalar-tensor gravity given by the action Eq.~(\ref{eq:jordanaction}) and scalar field potential Eq.~(\ref{eq:scalpot}).
We first describe the linear growth of structure for $\Lambda$CDM models and in the quasistatic regime of scalar-tensor gravity in \tsec{sec:lingrow}.
In \tsec{sec:sphcoll}, we discuss the spherical collapse model for chameleon theories and in \tsec{sec:env}, we examine the role of the environmental density in this approach.
Finally, in \tsec{sec:sims}, we give details on the $N$-body simulations employed in our study, which we use in \tsec{sec:halomodel} to test the chameleon spherical collapse and halo model predictions.


\subsection{Linear growth of structure in scalar-tensor theories} \label{sec:lingrow}

Combining the linearly perturbed Einstein field equations with the energy-momentum conservation in the total matter gauge of $\Lambda$CDM yields an ordinary second-order differential equation for the evolution of the matter overdensity $\Delta_{\rm m}(a,k)$,
\bq
 \Delta_{\rm m}'' + \left[ 2 - \frac{3}{2}\Om(a) \right] \Delta_{\rm m}' - \frac{3}{2}\Om(a) \Delta_{\rm m} = 0, \label{eq:lingroweq}
\eq
where we have defined $\Om(a) \equiv H_0^2\Om a^{-3}/H^2$ with the matter energy density parameter $\Om$ and the Hubble parameter and constant $H$ and $H_0$, respectively.
We replace $\Delta_{\rm m}$ in Eq.~(\ref{eq:lingroweq}) with the linear growth function
\bq
 D(a) \equiv \frac{\Delta_{\rm m}(a,k)}{\Delta_{\rm m}(a_{\rm i},k)}D(a_{\rm i}) \label{eq:lingrow}
\eq
and solve for $D(a)$ assuming matter domination with the corresponding initial conditions $D(a_{\rm i})=a_{\rm i}$ and $D'(a_{\rm i})=a_{\rm i}$ at an initial scale factor $a_{\rm i} \ll 1$.
Here and throughout the paper, $D(a)$ shall refer to the linear growth function of a $\Lambda$CDM cosmology.

In scalar-tensor theories, Eq.~(\ref{eq:lingrow}) is modified and in the quasistatic limit reads~\cite{esposito:00,tsujikawa:08a,lombriser:13a}
\bq
 \Delta_{\rm m}'' + \left[ 2 - \frac{3}{2}\Om(a) \right] \Delta_{\rm m}' - \frac{3}{2\bscal}\left[1+\frac{1}{3+2\omega}\frac{k^2\bscal}{a^2 m^2 + k^2\bscal}\right]\Om(a)\Delta_{\rm m} \simeq 0, \label{eq:modlingroweq}
\eq
describing the time and scale dependent linear growth function $D_{\scal}(a,k)$ defined as in Eq.~(\ref{eq:lingrow}) for scalar-tensor theories.
Note that contrary to $\Lambda$CDM, at near-horizon scales, the quasistatic $D_{\scal}(a,k)$ obtained from solving Eq.~(\ref{eq:modlingroweq}) deviates from the growth of matter fluctuations inferred from solving the full linear cosmological perturbation theory.
For the scalar-tensor theories of interest here, however, these deviations are small~\cite{lombriser:10,hojjati:12,lombriser:13a,lima:13,noller:13} and can safely be neglected in the high-curvature regime studied in the following (cf.~\cite{lombriser:11a,lombriser:13a}).
Moreover, we set $\bscal\simeq1$ in Eq.~(\ref{eq:modlingroweq}).

The linear matter power spectrum for the chameleon models $P_{{\rm L}\scal}$ can be determined from rescaling the $\Lambda$CDM power spectrum $P_{{\rm L}\Lambda{\rm CDM}}$,
\bq
 P_{{\rm L}\scal}(a,k) = \left( \frac{D_{\scal}(a,k)}{D(a)} \right)^2 P_{{\rm L}\Lambda{\rm CDM}}(a,k),
\eq
where we assume the same initial conditions for the scalar field models as in $\Lambda$CDM (see \tsec{sec:sims}).
We define the variance by evolving the initial matter fluctuations according to $\Lambda$CDM,
\bq
 S(a,r) \equiv \sigma^2(a,r) = \int  \rmd^3\mathbf{k} \, |\tilde{W}(k\,r)|^2 P_{{\rm L}\Lambda{\rm CDM}}(a,k)
= \frac{D^2(a)}{D^2(a_{\rm i})} \int  \rmd^3\mathbf{k} \, |\tilde{W}(k\,r)|^2 P_{\rm i}(a_{\rm i},k), \label{eq:variance}
\eq
where $\tilde{W}(k\,r)$ is a window function, obtained by the Fourier transform of a top-hat function of radius $r$, and $P_{\rm i}$ is the initial power spectrum at time $a_{\rm i}$.
We also apply this definition to chameleon models, i.e., using $D(a)$ in the extrapolation of the initial matter power spectrum rather than $D_{\scal}(a,k)$.
In this case, the variance has to be interpreted as an effective quantity.
We discuss the advantages of this approach in \tsec{sec:massfctbias}.
Note that we can also write $S$ as a function of the mass $M=4\pi\,\rhomb\,r^3/3$ enclosed by the top-hat function instead of $r$.


\subsection{Chameleon spherical collapse} \label{sec:sphcoll}

The formation of clusters can be studied with the spherical collapse model, where the dark matter halo is approximated by a spherically symmetric top-hat overdensity, which is evolved according to the nonlinear continuity and Euler equations from an initial time to the time of its collapse.
The chameleon suppression in the spherical collapse calculation can be incorporated in this model following Ref.~\cite{li:11a} (cf.~\cite{borisov:11}), who allow a mass and environment dependent modification of the gravitational force by implementing the thin-shell thickness estimator for the chameleon transition by Ref.~\cite{khoury:03b} described in \tsec{sec:intermediate}.
The chameleon spherical collapse model has also been adapted to $f(R)$ gravity and applied in the description of halo mass functions produced by $N$-body simulations thereof, yielding good agreement between the two~\cite{lombriser:13b}.
We generalise this description to chameleon models of the form given by Eqs.~(\ref{eq:jordanaction}) and (\ref{eq:scalpot}), and in \tsec{sec:halomodel}, study its application in the modelling of halo properties.

In the following discussion, we denote the physical radius of the top-hat overdensity at time $a$ by $\zeta(a)$, which at the initial time $a_{\rm i}\ll1$ is determined by $\zeta(a_{\rm i})=a_{\rm i}\RTH$.
Due to the nonlinear evolution of the overdensity, $\zeta(a)$ deviates from this simple linear relation when $a>a_{\rm i}$, which can be described by defining the dimensionless variable $y\equiv \zeta(a)/a\RTH$ with $\tilde{\rho} = \rhom/\rhomb = y^{-3}$ as $\rhomb a^3 \RTH^3 = \rhom \zeta^3$ due to conservation of mass enclosed in the overdensity.
The evolution of the physical radius of the spherical shell is governed by the equation of motion~\cite{schmidt:08,li:11a,lombriser:13b}
\bq
 \frac{\ddot{\zeta}}{\zeta} \simeq -\frac{\kappa^2}{6} \left( \rhomb - 2\bar{\rho}_{\Lambda} \right) - \frac{\kappa^2}{6} \left(1 + \frac{\Delta F}{F_{\rm N}} \right) \drhom,
 \label{eq:shellmotion}
\eq
where dots denote cosmic time derivatives and we use the gravitational force modification $\Delta F/F_{\rm N}$ given in Eq.~(\ref{eq:enhf}), replacing $\Delta r / \RTH \rightarrow \Delta\zeta/\zeta$.
Thus, with $\tilde{\rho}_{\rm in}=\yhal^{-3}$, we obtain the evolution of $\yhal$ from solving
\bq
 \yhal'' + \left[ 2 - \frac{3}{2} \Om(a) \right] \yhal' + \frac{1}{2} \Om(a) \left(1 + \frac{\Delta F}{F_{\rm N}} \right) \left( \yhal^{-3} - 1 \right) \yhal = 0 \label{eq:yhal}
\eq
with the force enhancement following from Eq.~(\ref{eq:thsh}),
\bq
 \frac{\Delta \zeta}{\zeta} \simeq \frac{(3+2\omega)(\bscal_0-1) \, a^{\frac{4-\alpha}{1-\alpha}}}{3\Om(H_0\RTH)^2} \yhal \left[ \left( \frac{1+4\frac{\Olam}{\Om}}{\yhal^{-3} + 4\frac{\Olam}{\Om} a^3} \right)^{\frac{1}{1-\alpha}} - \left( \frac{1+4\frac{\Olam}{\Om}}{\yenv^{-3} + 4\frac{\Olam}{\Om} a^3} \right)^{\frac{1}{1-\alpha}} \right]. \label{eq:thshsphcoll}
\eq
For the environment $\tilde{\rho}_{\rm out}=\yenv^{-3}$, we assume a $\Lambda$CDM evolution, obtained from Eq.~(\ref{eq:shellmotion}) in the limit $\Delta\zeta/\zeta\rightarrow0$ or equivalently, $\Delta F\rightarrow0$, which yields
\bq
 \yenv'' + \left[ 2 - \frac{3}{2} \Om(a) \right] \yenv' + \frac{1}{2} \Om(a) \left( \yenv^{-3}-1 \right) \yenv = 0, \label{eq:yenv}
\eq
forming a system of differential equations together with Eq.~(\ref{eq:yhal}).
We solve this system, setting the initial conditions in the matter-dominated regime,
\bq
 y_{\rm h/env, i} = 1 - \frac{\delta_{\rm h/env, i}}{3}, \ \ \ \ \ y_{\rm h/env, i}' = - \frac{\delta_{\rm h/env, i}}{3},
\eq
at an initial scale factor $a_{\rm i} \ll 1$.
Finally, analogous to the description of the variance in \tsec{sec:lingrow}, we  define an effective linear overdensity for the chameleon theories,
\bq
 \delta_{\rm h/env}({\bf x}; \zeta_{\rm h/env}) \equiv \frac{D(a)}{D(a_{\rm i})} \delta_{\rm h/env, i}, \label{eq:extrapolation}
\eq
extrapolating initial overdensities to time $a$ using the $\Lambda$CDM linear growth function $D(a)$, which is obtained from solving Eq.~(\ref{eq:lingroweq}) with the definition Eq.~(\ref{eq:lingrow}).
In particular, we use Eq.~(\ref{eq:extrapolation}) to define the linear collapse density $\deltac$ and the environmental density $\denv$, corresponding to the effective linear overdensity at the time when Eq.~(\ref{eq:yhal}) produces a singularity given $\delta_{\rm h, i}$ and the associated linear extrapolation of $\delta_{\rm env, i}$, respectively.


\subsection{Environmental density distribution} \label{sec:env}

\begin{figure}
 \centering
 \resizebox{0.5\hsize}{!}{\includegraphics{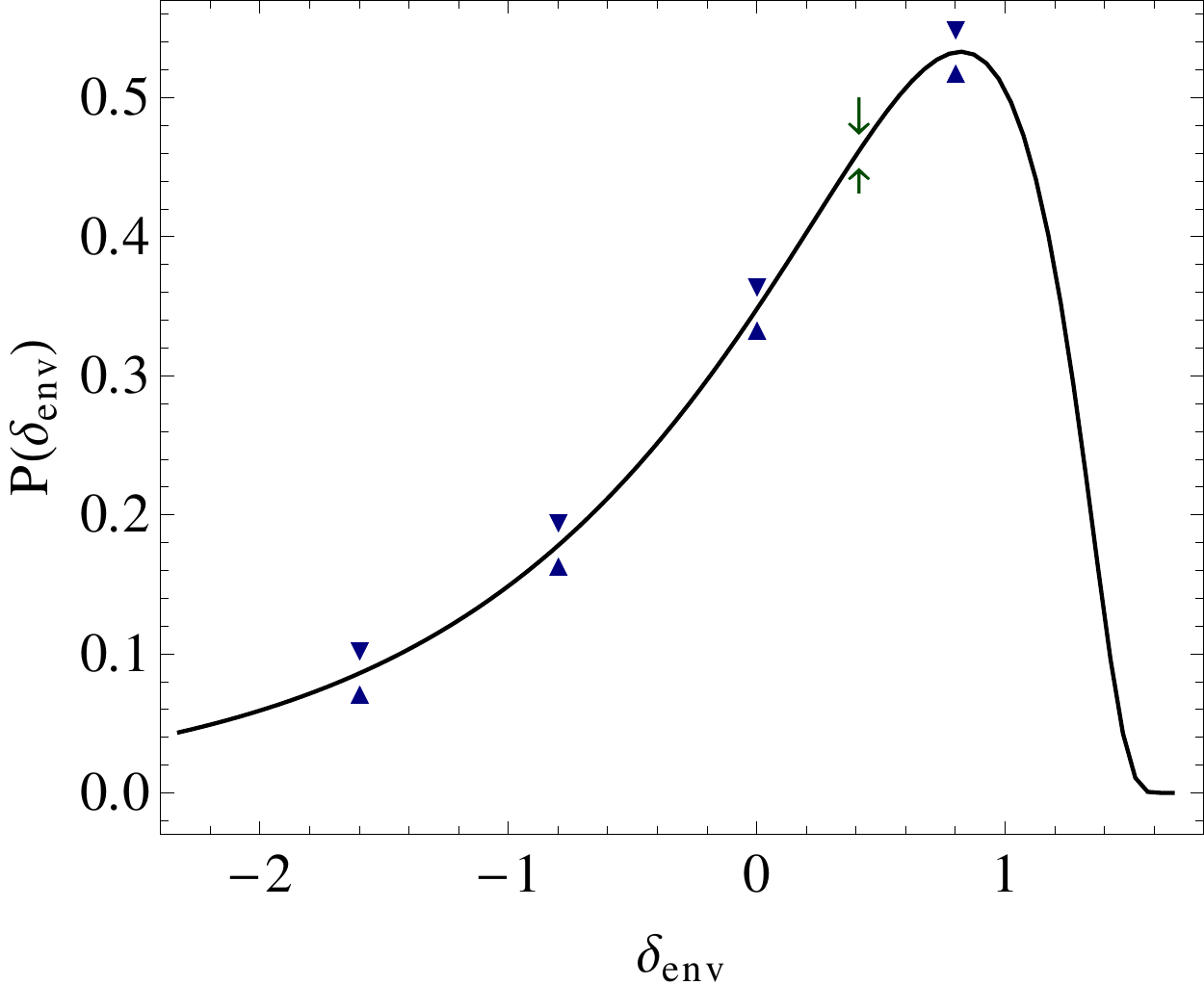}}
 \caption{
  Probability distribution $P(\denv)$ of the Eulerian environment.
The blue diamonds indicate the different environmental densities studied in \tsec{sec:halomodel} and the green arrows mark the environmental density for which the collapse density approximately matches the environmentally averaged counterpart $\langle\deltac\rangle_{\rm env}$.
  }
\label{fig:env}
\end{figure}

The environmental density $\denv$ or $\delta_{\rm env,i}$ affects the gravitational force enhancement via Eq.~(\ref{eq:thshsphcoll}) and thus, the enhancement in the growth of structure due to this modification.
In order to correctly reproduce the halo properties of chameleon theories measured in $N$-body simulations and to perform consistent tests of the gravitational modifications against observations, we need to incorporate the environmental effects in our modelling.
This can be done by either characterising the quantities measured in simulations and from observations by their different environmental densities and comparing this to the according predictions from the spherical collapse model or by studying the averaged observables against the environmentally averaged spherical collapse predictions.

We follow the second approach and define the size of the environments by their Eulerian (physical) radius $\zeta$ and therefor adopt the value $\zeta=5h^{-1}~{\rm Mpc}$ used in Refs.~\cite{li:12b,lombriser:13b}.
The probability distribution of the Eulerian environmental density $\denv$ can approximately be described by~\cite{lam:08,li:12b,lam:12}
\bq
 P_{\zeta}(\denv) = \frac{\beta^{\varpi/2}}{\sqrt{2\pi}} \left[ 1 + (\varpi - 1)\frac{\denv}{\deltac} \right] \left( 1 - \frac{\denv}{\deltac} \right)^{-\varpi/2-1} \exp \left[ -\frac{\beta^{\varpi}}{2} \frac{\denv^2}{(1 - \denv/\deltac)^{\varpi}} \right] \label{eq:eulerian}
\eq
with $\beta = (\zeta/8)^{3/\deltac} / \sigma_8^{2/\varpi}$, $\varpi = \deltac \gamma$, and
\bq
 \gamma = - \frac{\rmd \ln S_{\xi}}{\rmd \ln M_{\rm env}} = \frac{\tilde{n}_{\rm s}+3}{3},
\eq
where we assume that the environment evolves according to $\Lambda$CDM.
Here, $\xi$ is the Lagrangian (or initial comoving) radius
with $\xi=8h^{-1}~{\rm Mpc}$ such that $S_{\xi}=\sigma_8^2$ and $\tilde{n}_{\rm s}$ is the slope of the matter power spectrum $P_{\rm L}(k)$ on large scales at an initial time $a_{\rm i}\ll1$ in the matter era after the turn over, corresponding to the initial time used in the excursion set calculation.

We show the distribution $P_{\zeta}(\denv)$, assuming cosmological parameter values as defined in \tsec{sec:sims} in Fig.~\ref{fig:env}.
It can be used to determine the environmentally averaged linear collapse density $\langle\deltac\rangle_{\rm env}$, which can then be applied in the modelling of halo properties.
We will study this case along with further environmental densities, $\denv\simeq0.8,0,-0.8,-1.6$, corresponding to the locations of the diamond markers illustrated in Fig.~\ref{fig:env}.
Alternatively and more accurately, observables may first be modelled using $\deltac(\denv)$ with subsequent averaging over $P_{\zeta}(\denv)$.
The two approaches yield only small deviations in the corresponding halo mass functions~\cite{lombriser:13b} such that here, for simplicity, we shall use the first method.
Note that in two further, very simple, and approximately model-independent approaches, one can approximate the averaged observable by evaluating it at the peak of the environmental distribution $\denv\approx0.8$ or for the average environment $\langle\denv\rangle_{\rm env}\approx0.16$ instead, for which deviations from using $\langle\deltac\rangle_{\rm env}\approx\deltac(\denv=0.4)$ can be estimated by comparing the results obtained for the different environmental densities shown in Fig.~\ref{fig:env}. 

We refer to Refs.~\cite{li:12b,lam:12,lombriser:13b} for more details on the role of the environment in determining the chameleon modifications.


\subsection{$N$-body simulations of chameleon $f(R)$ gravity} \label{sec:sims}

Finally, $N$-body simulations are an essential tool for understanding the formation of large-scale structure on nonlinear scales and as such provide a great laboratory for studying the chameleon mechanism.
Thus, in order to test the accuracy of our predictions obtained from the chameleon spherical collapse and halo model in \tsec{sec:halomodel}, we restrict to cases with $\omega=0$ and use the large-volume, high-resolution dark matter chameleon $f(R)$ gravity simulation output of Ref.~\cite{li:12c}.
These simulations are performed using an adaptive particle mesh code~\cite{li:11}, which solves the quasistatic modified Poisson and scalar field equations (see, e.g.,~\cite{oyaizu:08a}), and cover the Newtonian and chameleon scenarios for each field strength $|\bscal_0-1|=|\bar{f}_{\bar{R}0}|=10^{-6}, 10^{-5}, 10^{-4}$ with exponent $\alpha=1/2$ ($n=1$).
The cosmological parameters are set to $\Om=1-\Olam$ with $\Olam=0.76$, $h=0.73$ for the dimensionless Hubble constant, $n_{\rm s}=0.958$ is the slope of the primordial power spectrum, and the initial power in curvature fluctuations $A_{\rm s}$ is set to correspond to a power spectrum normalisation $\sigma_8\equiv\sigma(a=1,r=8~h^{-1}{\rm Mpc})=0.8$ in $\Lambda$CDM.
We use a simulation of box size $L_{\rm box} = 1.0~\hGpc$ and total particle number $N_{\rm p}=1024^3$.
The grid structure efficiently follows the density distribution to better resolve the high-density regions as during the simulation, in regions where the local densities are sufficiently large to reach a predefined threshold, the domain grids are progressively refined.
We use a spherical overdensity algorithm~\cite{jenkins:00} to identify halos within the simulation and their associated masses.
In the process of defining the halos, we use the virial overdensity $\Delta_{\rm vir}\approx390$, which is obtained assuming $\Lambda$CDM.
We also apply this value to identify halos produced in scalar-tensor gravity, which allows us to make a fair comparison between the different models based on an equal-overdensity approach.
Note that the error of using $\Lambda$CDM virial masses $\Mvir$ instead of accurate virial masses for the chameleon models is estimated to be small compared to the overall modification from the enhanced force~\cite{lombriser:13b}.
Furthermore, note that we restrict to quasistatic $f(R)$ gravity $N$-body simulations, where deviations from relaxing this assumption are expected to be small~\cite{oyaizu:08a,llinares:13b}.


\section{Chameleon halo modelling} \label{sec:halomodel}

Using the chameleon spherical collapse model described in \tsec{sec:sphcoll}, we model the halo mass function and the linear halo bias in the peak-background split using the Sheth-Tormen prescription~\cite{sheth:99} in \tsec{sec:massfctbias}.
In \tsec{sec:concprof}, we then approximate the chameleon halos by NFW profiles and model the characteristic density and scale via the halo concentration, for which we introduce a scaling function based on a fit to $\Lambda$CDM $N$-body simulations.
Finally, in \tsec{sec:pmm}, we combine these descriptions to approximate the nonlinear matter power spectrum using the halo model.


\subsection{Halo mass function and linear halo bias} \label{sec:massfctbias}

The statistics of virialised clusters can be described using excursion set theory, where collapsed structures correspond to regions for which the smoothed initial matter density fields exceed the threshold given by the collapse density $\deltac$.
The size of such a region relates to the variance $S$ via the integration of the power spectrum $P_{\rm L}(k)$ in Eq.~(\ref{eq:variance}).
If the wavenumbers are uncorrelated, an incremental step in the smoothed initial overdensity field from changing $S$ is independent of its previous values.
The smoothed matter density field is then described by a Brownian motion in $S$ with Gaussian probability distribution, for which the increment is a Gaussian field with zero mean.
The Press-Schechter~\cite{press:74} expression describes the distribution $f$ of the Brownian motion trajectories that first cross the flat barrier $\deltac$ at $S$.

\begin{figure}
 \centering
 \resizebox{\hsize}{!}{
 \resizebox{0.4975\hsize}{!}{\includegraphics{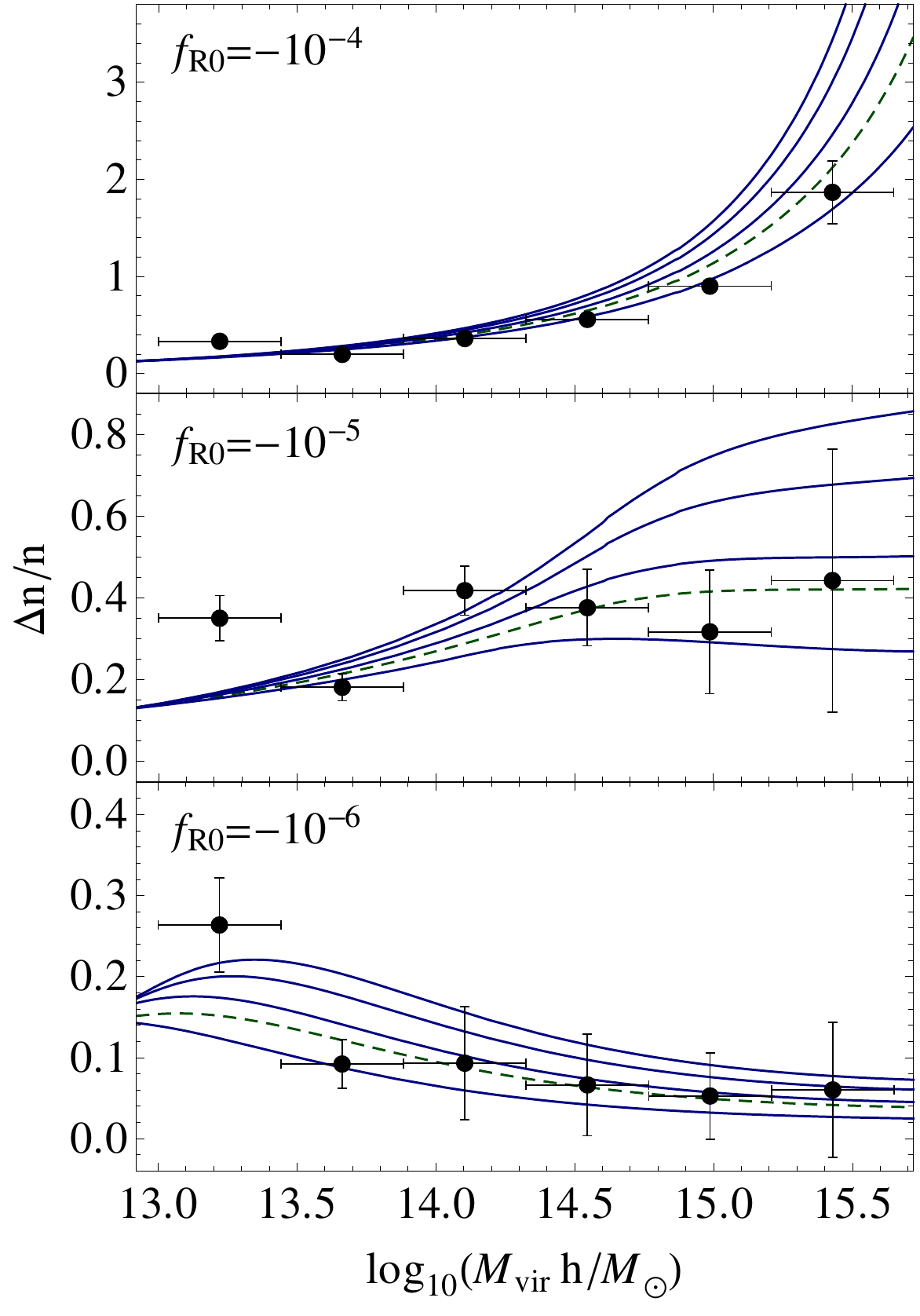}}
 \resizebox{0.5025\hsize}{!}{\includegraphics{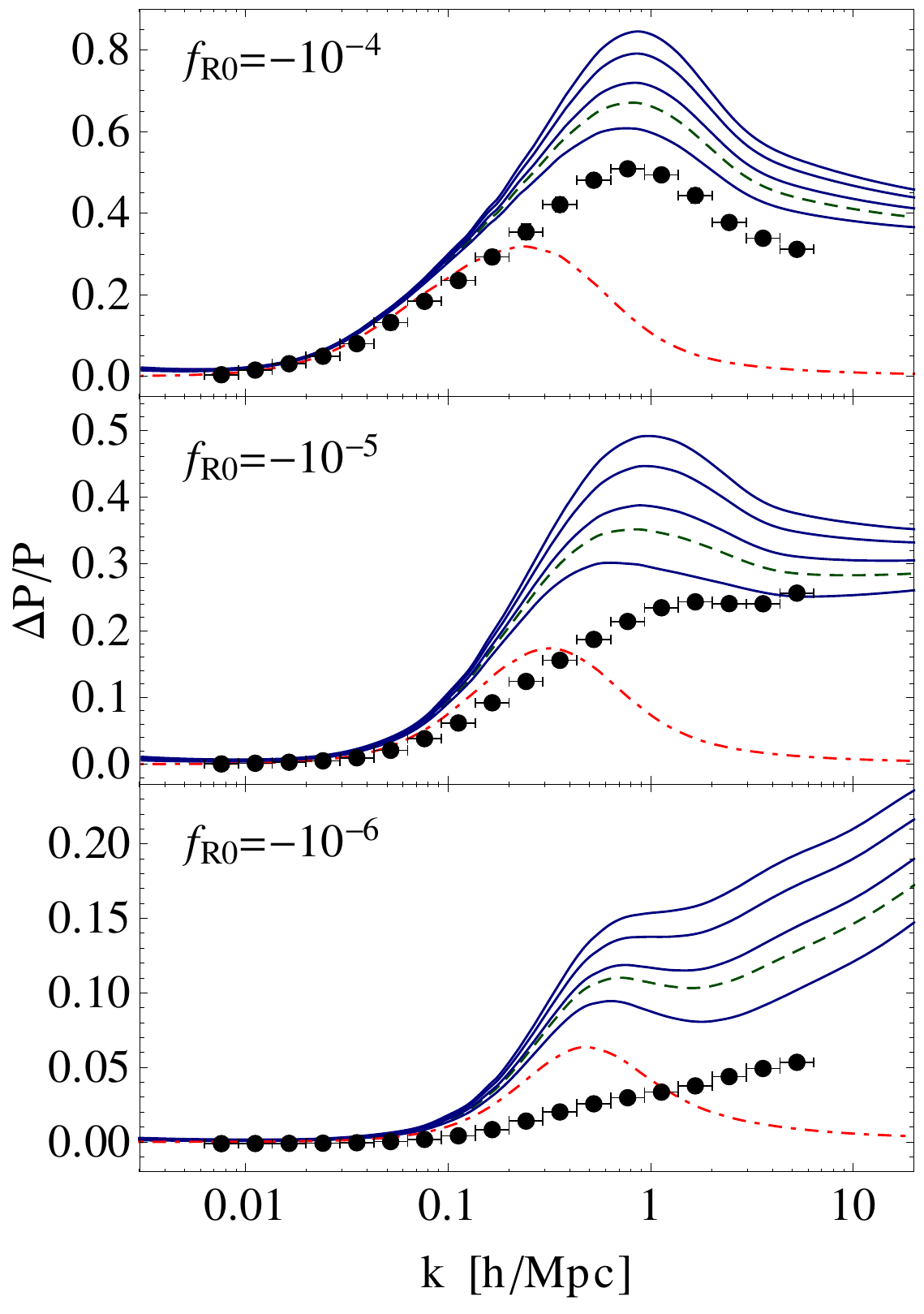}}
 }
 \caption{
  Relative difference between chameleon $f(R)$ and $\Lambda$CDM halo mass function $n_{\ln\Mvir}$ (left panel) and nonlinear matter power spectrum $P(k)$ (right panel) determined from the spherical collapse model, the Sheth-Tormen prescription, the NFW halo profile, and the halo model.
  The blue solid lines indicate the enhancements obtained for the choices of different environmental densities shown in Fig.~\ref{fig:env}.
  The green dashed and red dot-dashed curves correspond to the enhancements obtained from the environmental density $\denv$ for which $\deltac(\denv)\approx\langle\deltac\rangle_{\rm env}$ and from adopting halo properties for the $f(R)$ chameleon model that are equivalent to their counterparts from $\Lambda$CDM, respectively.
  We compare our predictions against results from the $N$-body simulations (data points) of Ref.~\cite{li:12c}.
  }
\label{fig:massfctpower}
\end{figure}

For chameleon models, however, due to the scale-dependent modification of gravity, the barrier is no longer flat and depends on $S$ and the environment embedding the collapsing halo.
Similarly, if relaxing the assumption of sphericity of the halo, the barrier becomes dependent on $S$.
Motivated by excursion set theory with a moving barrier such as caused by ellipsoidal collapse~\cite{sheth:99b, sheth:01}, Sheth and Tormen~\cite{sheth:99} introduced a modification of the Press-Schechter expression for the first-crossing distribution $f$ given by
\bq
 \nu \, f(\nu) = \mathcal{N} \sqrt{\frac{2}{\pi} q \, \nu^2} \left[ 1 + \left(q \, \nu^2\right)^{-p} \right] e^{-q\,\nu^2/2}, \label{eq:st}
\eq
where $\nu\equiv\deltac/\sqrt{S}$ is the peak-threshold, $\mathcal{N}$ is a normalisation parameter such that $\int \rmd \nu \, f(\nu) = 1$, $p=0.3$, and $q=0.707$ is set to match results from $\Lambda$CDM $N$-body simulations in the halo mass function
\bq
 n_{\ln\Mvir} \equiv \frac{\rmd n}{\rmd\ln\Mvir} = \frac{\rhomb}{\Mvir} f(\nu) \frac{\rmd \nu}{\rmd\ln\Mvir}.
\eq
Eq.~(\ref{eq:st}) has been shown to also provide good fits to $N$-body simulations of linearised $f(R)$ gravity~\cite{schmidt:08} and chameleon $f(R)$ gravity~\cite{lombriser:13b} when using the mass and environment dependent collapse density described in \tsec{sec:sphcoll}.
We adopt this approach to model the halo mass function of the scalar-tensor models given by Eqs.~(\ref{eq:jordanaction}) and (\ref{eq:scalpot}), using the effective linear collapse density $\deltac$ obtained from the chameleon spherical collapse and the effective variance in Eq.~(\ref{eq:variance}) to determine the chameleon peak-threshold $\nu$ (cf.~\cite{li:11b,kopp:13}).
Note that the effective collapse density $\deltac$ is defined by the extrapolation of the initial overdensity leading to collapse at $a$  using the $\Lambda$CDM growth of structure $D(a)$ from Eq.~(\ref{eq:lingroweq}).
Similarly, the variance $S$ in Eq.~(\ref{eq:variance}) is obtained from the integration of the initial matter power spectrum and extrapolated to $a$ using $D(a)$.
Hence, we have $\nu=\deltac/\sqrt{S}=\delta_{c,\rm i}/\sqrt{S_{\rm i}}=\nu_i$ due to the scale-independent growth of structure in $\Lambda$CDM.
In contrast, using $D_{\scal}(a,k)$ for the extrapolation of $\delta_{c,\rm i}$ and $S_{\rm i}$, we have $\nu\neq\nu_{\rm i}$ in general.
Thus, our definition corresponds to defining the peak-threshold at initial time.
We compare the Sheth-Tormen halo mass function using this peak-threshold with $N$-body simulations of $f(R)$ gravity in Fig.~\ref{fig:massfctpower}, finding good agreement between the two.
Note that the halo-finder employed does not identify and remove subhalos from the $N$-body simulations, which leads to a contamination of the signature at the low-mass end.

The nonlinear dark matter distribution is biased with respect to the linear distribution and in order to determine this bias, on large scales, one can perform a peak-background split.
In this approach, short-wavelength modes are regarded as superimposing the long-wavelength modes, perturbing $\deltac$ and modifying the collapse.
This perturbation can be expressed via the halo mass function and for the Sheth-Tormen expression Eq.~(\ref{eq:st}) yields the linear halo bias~\cite{sheth:99}
\bq
 b_{\rm L}(\Mvir) \equiv b(k=0,\Mvir) = 1 + \frac{a\,\nu^2-1}{\deltac} + \frac{2p}{ \deltac \left[ 1 + \left( a\,\nu^2 \right)^p \right] }. \label{eq:blin}
\eq
In chameleon models, with the relative suppression of the effective linear collapse density $\deltac$ with respect to $\Lambda$CDM, $b_{\rm L}$ decreases.
Using the spherical collapse model described in \tsec{sec:sphcoll} to compute $\deltac$, this modification becomes mass and environment dependent.
Note that the halo bias only marginally affects the halo model computation of the power spectrum in \tsec{sec:pmm} through the two-halo term and we do not show it here.
We refer the reader to Ref.~\cite{schmidt:08} for a measurement of the linear halo bias from $N$-body simulations of $f(R)$ gravity.


\subsection{Halo concentration and halo density profiles} \label{sec:concprof}

In \tsec{sec:theory}, we have assumed that chameleon halos are spherically symmetric and that their radial density profiles are well described by the NFW fitting function Eq.~(\ref{eq:nfw}).
The characteristic density and scale of the NFW fit can be modelled using Eqs.~(\ref{eq:chardens}) and (\ref{eq:charscale}), respectively, which are functions of the virial mass and concentration, $\Mvir$ and $\cvir$, where we take $\Delta_{\rm vir}=390$ as in \tsec{sec:sims}.
We can reduce this dependency to a function of mass only by adopting a mass-concentration scaling relation such as $\cvir(\Mvir,a) = 9 a (\Mvir/M_*)^{-0.13}$, which has been calibrated to $\Lambda$CDM $N$-body simulations in Ref.~\cite{bullock:99} using approximately $5\times10^3$ halos of mass $10^{11}-10^{14}~\Msunh$.
Hereby, the critical mass $M_*$ satisfies $S(M_*)=\deltac^2$.
Note that we assume that the applicability of this calibration can be extended to more massive halos.
This scaling relation can further be applied to estimate the concentration of halos formed in scalar-tensor gravity.
This approach was taken for $f(R)$ gravity ($\omega=0$) in Refs.~\cite{schmidt:08,lombriser:11b,li:11b}, in which $M_*$ is determined by solving $S_{\scal}(M_*)=\delta_{{\rm c}\scal}^2$, where $\delta_{{\rm c}\scal}$ is given by the spherical collapse in the limiting cases of either a $\Delta F/F_{\rm N}=(3+2\omega)^{-1}$ or a $\Delta F=0$ modification and $S_{\scal}$ is the true variance of the scalar-tensor model, i.e., from using $D_{\scal}(a,k)$ instead of $D(a)$ in Eq.~(\ref{eq:variance}).
We have followed this approach, generalised to non-zero values of $\omega$, in \tsec{sec:locconst} to obtain a simple estimation for the concentration of the Milky Way halo given its measured mass and derive approximate constraints on the scalar field amplitude $|\bscal_0-1|$.

As was pointed out in Ref.~\cite{lombriser:12}, this approach does, however, not incorporate a chameleon screening effect; $M_*$ is determined from a given scalar field amplitude $(\bscal_0-1)$ and only introduces a constant shift of the concentration, independent of mass and environment.
Here, we evade this deficiency by reinterpreting the mass-concentration relation.
We apply the inverse function of the variance to $\sigma(M_*)=\deltac$ to define the critical mass as $M_*(\deltac,\sigma)\equiv\sigma^{-1}\circ\deltac$.
This corresponds to assigning an effective flat barrier at each mass and environmental density bin $(M_{{\rm vir},j},\delta_{{\rm env},k})$ with $\delta_{{\rm c},jk}\equiv\deltac(M_{{\rm vir},j},\delta_{{\rm env},k})$, interpreting this as a $\Lambda$CDM threshold, and performing the standard computation of $M_{*,jk}$ via $\delta_{{\rm c},jk}=\sigma(M_{*,jk})$.
Hence, varying $\deltac$ as a function of mass and environment, where $\sigma$ is determined from the mass, the concentration becomes
\bqa
 \cvir(\Mvir,\denv,a) & = & 9 a \left[ \frac{M_*(\Mvir,\denv)}{\Mvir} \right]^{0.13}, \label{eq:concentration} \\
 M_*(\Mvir,\denv) & \equiv & (\sigma^{-1}\circ\deltac)(\Mvir,\denv) = \sigma^{-1}(\deltac(\Mvir,\denv)), \label{eq:massstar}
\eqa
which introduces a chameleon screening effect in $\cvir$.

Given the halo concentration, we can compute the characteristic density $\rhos$ and characteristic radius $\rs$ of the NFW profile through Eqs.~(\ref{eq:chardens}) and (\ref{eq:charscale}).
While $\rhos$ is enhanced due to the presence of the chameleon field, $\rs$ becomes smaller compared to its $\Lambda$CDM counterpart.
Thereby, the mass and environmental dependence of the halo concentration is reflected in the modifications of $\rhos$ and $\rs$.
Note that the halo concentration and, hence, $\rhos$ and $\rs$, only marginally affect the halo model computation of the power spectrum in \tsec{sec:pmm} at the smallest scales, mainly through the one-halo term, and we do not show them here.
We refer the reader to Ref.~\cite{lombriser:12} for a measurement of the halo concentration, characteristic density, and characteristic radius from $N$-body simulations of $f(R)$ gravity.


\subsection{Nonlinear matter power spectrum} \label{sec:pmm}

Finally, we use the halo model~\cite{peacock:00,seljak:00,cooray:02} to decompose statistics of cosmological structures into the underlying halo contributions.
In this picture, the nonlinear matter power spectrum can be described by the two-halo and one-halo terms,
\bqa
 P_{\rm mm}(k) & \simeq & I^2(k) P_{\rm L}(k) + P^{1h}(k), \label{eq:nlpmm} \\
 P^{1h}(k) & = & \int \rmd\ln\Mvir n_{\ln\Mvir} \frac{\Mvir^2}{\rhomb^2} \left| y(k,\Mvir) \right|^2 \label{eq:p1hk}
\eqa
with
\bq
 I(k) \simeq \int \rmd\ln\Mvir n_{\ln\Mvir} \frac{\Mvir}{\rhomb} y(k,\Mvir) b_{\rm L}(\Mvir), \label{eq:ik}
\eq
where $y(k,M)$ shall be the Fourier transform of a NFW density profile which is truncated at $\rvir$ and normalised as $\lim_{k\rightarrow0} y(k,M) = 1$.
We further require $\lim_{k\rightarrow0} I(k) = 1$.
The expressions Eqs.~(\ref{eq:nlpmm}), (\ref{eq:p1hk}), and (\ref{eq:ik}) apply to both $\Lambda$CDM and the chameleon model.
The halo mass function, halo density profile, and halo bias are computed according to \tsecs{sec:massfctbias} and {\ref{sec:concprof}}.
The linear matter power spectrum for $\Lambda$CDM and the chameleon model, $P_{{\rm L}\Lambda{\rm CDM}}$ and $P_{{\rm L}\scal}$ described in \tsec{sec:lingrow}, respectively, are determined from the initial power spectrum using the Eisenstein-Hu transfer function~\cite{eisenstein:97a,eisenstein:97b}.

We compare the nonlinear matter power spectrum predicted by the halo model in Eq.~(\ref{eq:nlpmm}) with the power spectrum obtained from $N$-body simulations for chameleon $f(R)$ gravity in Fig.~\ref{fig:massfctpower}.
The blue solid curves illustrate the effects of assuming the different environmental densities of \tsec{sec:env} when determining the halo properties used
in Eqs.~(\ref{eq:p1hk}) and (\ref{eq:ik}).
They mainly contribute via the one-halo and to a smaller extent through the two-halo contribution.
The effect of the environment can be interpreted as an average over unscreened, screened, and partially screened forces between the dark matter particles.
The green dashed curve corresponds to the average obtained for the environmental density $\denv$ for which $\langle\deltac\rangle_{\denv}\approx\deltac(\denv)$ and
the red dot-dashed curve represents the case in which $I(k)$ and the one-halo term correspond to the contributions expected in a $\Lambda$CDM model, or equivalently, where the halo mass function, halo bias, and halo profile are computed for an extreme high-density environment.

Compared to $\Lambda$CDM, in $f(R)$ gravity, nonlinearities contribute at slightly larger scales; whereas for $\Lambda$CDM, at $k=0.1~\Mpch$, the linear power spectrum is about 5\% smaller than its nonlinear counterpart, at the same scale, in $f(R)$ gravity, for $|\bar{f}_{\bar{R}0}|=10^{-4}$, this deviation is about 10\%.
For both models, the two-halo term approximately corresponds to the linear power spectrum and only at small scales is suppressed with respect to $P_{\rm L}$, i.e., $\gtrsim5\%$ at $k\gtrsim1~\Mpch$.
It only marginally affects the interpolation between the linear power spectrum and the one-halo term.
In $\Lambda$CDM, the one-halo term starts to dominate over the two-halo term at $k\gtrsim(0.4-0.5)~\Mpch$.
This scale of equality between the two contributions is slightly shifted to larger scales for $f(R)$ models.

\begin{figure}
 \centering
 \resizebox{\hsize}{!}{
 \resizebox{0.499\hsize}{!}{\includegraphics{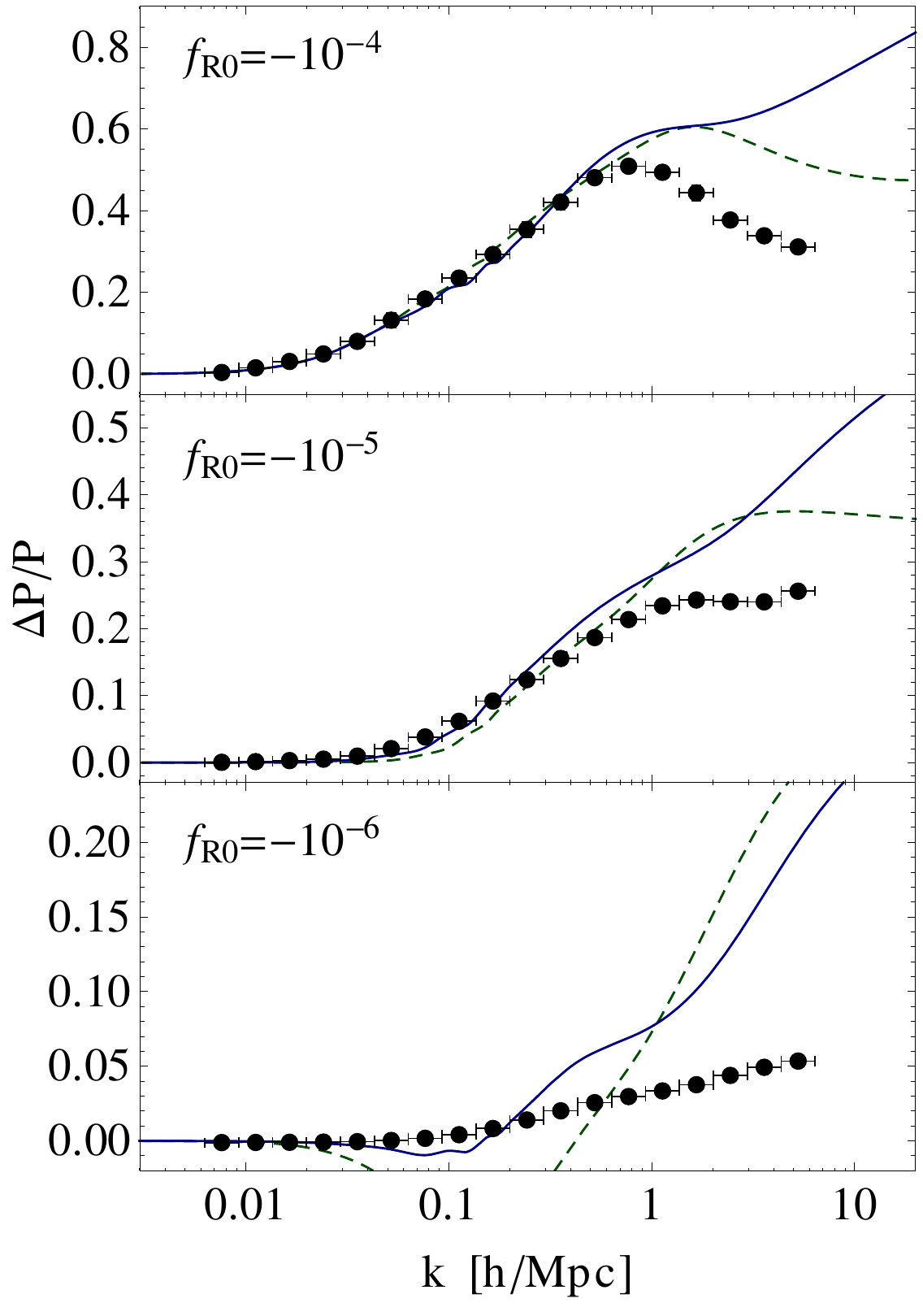}}
 \resizebox{0.501\hsize}{!}{\includegraphics{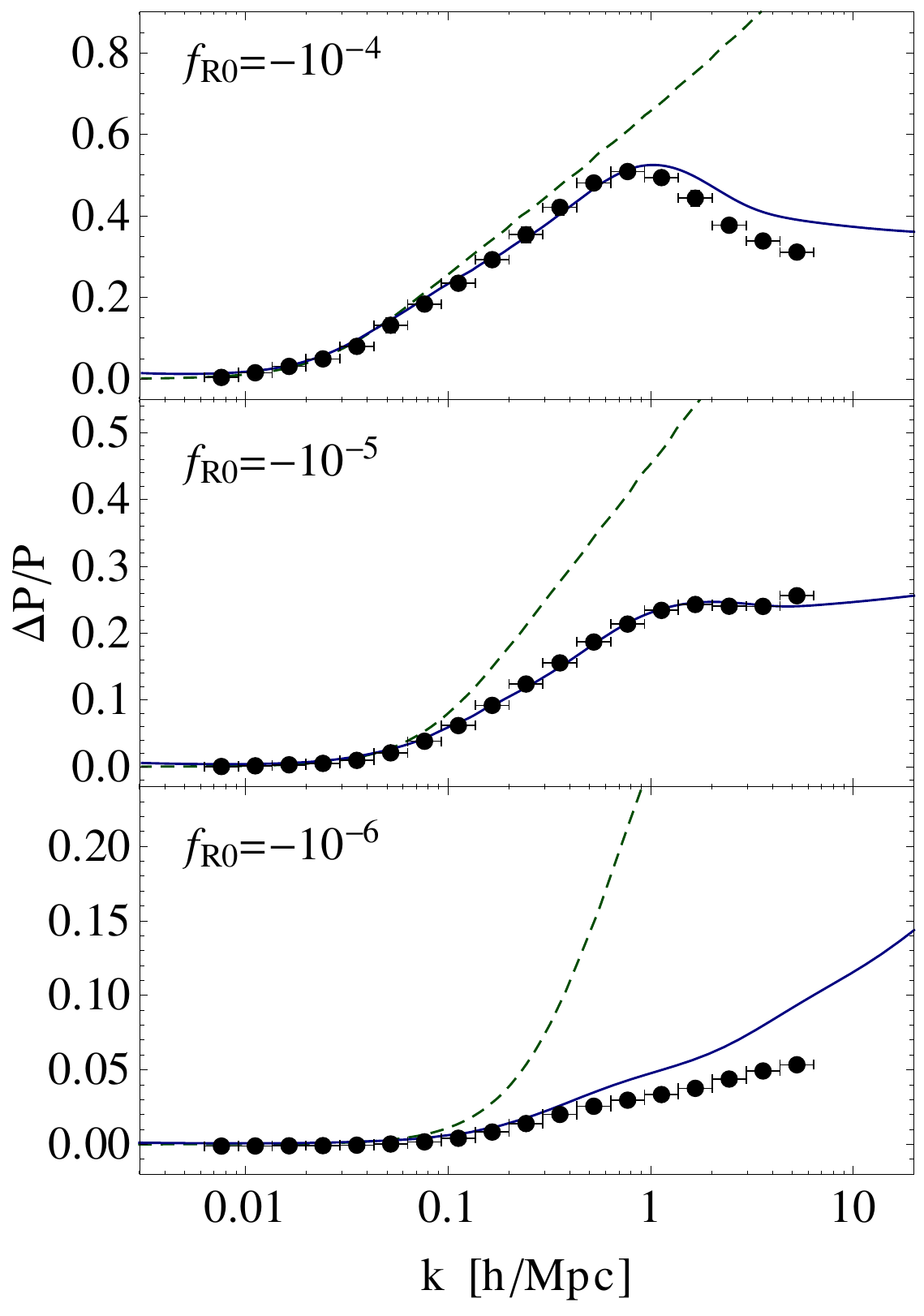}}
 }
 \caption{
  Same as Fig.~\ref{fig:massfctpower} but for different computations of the matter power spectrum.
  The different curves show the relative deviations between $f(R)$ and $\Lambda$CDM predictions obtained from using:
  the original~\cite{smith:02} (solid blue) and revised~\cite{takahashi:12} (dashed green) HALOFIT approaches in the left panel; 
  the halo model with modified two-halo term (solid blue) and the linear power spectra only (dashed green) in the right panel.
 }
\label{fig:pmm}
\end{figure}

In the left panel of Fig.~\ref{fig:pmm}, we show the relative enhancement of the power spectrum of $f(R)$ gravity with respect to $\Lambda$CDM predicted by the original~\cite{smith:02} and revised~\cite{takahashi:12} HALOFIT approaches.
Both descriptions fail to capture the suppression of the small-scale enhancement of the power in the small-field limit.
While the revised HALOFIT improves the description of the power spectrum enhancement for $|\bar{f}_{\bar{R}0}|=10^{-4}$ over the original version, it yields a worse fit to this enhancement in the chameleon screened regime $|\bar{f}_{\bar{R}0}|=10^{-6}$.
In comparison, the halo model provides a good qualitative description of the power spectrum at high $k$-modes but fails to reproduce the correct amplitude of the modification, particularly at intermediate scales, which are described by the two-halo term.
We use the linear power spectrum for the computation of this contribution, which underestimates nonlinear effects on these intermediate scales such as a chameleon suppression of the linearly computed growth enhancement.
To account for this deficiency, we introduce a simple interpolation function, replacing the linear power spectrum $P_{{\rm L}\scal}(k)$ with
\bq
 P_{{\rm L}\scal}^{\rm eff}(a,k) = \frac{P_{{\rm L}\scal}(a,k) + (k/k_*)P_{{\rm L}\Lambda{\rm CDM}}(a,k)}{1+k/k_*},
\eq
where $k_*=0.1\sqrt{(1-\bscal)/10^{-5}}~\Mpch$, motivated by the relation of scale and scalar field amplitude in Eq.~\ref{eq:thsh}.
For the computation of the one-halo term, we then assume the most probable environmental density, corresponding to the position of the peak of Fig.~\ref{fig:env}.
We show the resulting adjusted halo model prediction for the relative enhancement of the power spectrum in the right panel of Fig.~\ref{fig:pmm}, which is in good agreement with the $N$-body simulations over a wide range of scales.

Note that alternatively to our modified halo model approach, the description of the enhancement in the nonlinear matter power spectrum can be improved by employing fitting functions as have been devised in Refs.~\cite{hu:07b,zhao:10b,li:11b,zhao:13}, perturbation theory~\cite{koyama:09}, or a combination of the halo model with one-loop perturbations~\cite{brax:13b}.

\section{Conclusion} \label{sec:conclusion}

We generalise the Hu-Sawicki $f(R)$ gravity model to scalar-tensor models with constant Brans-Dicke parameter that match a $\Lambda$CDM expansion history and undergo chameleon screening of the scalar field and suppression of the gravitational modifications within high-density regions.
Based on Solar System constraints on possible deviations from GR, we formulate constraints on the model parameters of our scalar-tensor theories for being locally viable.

We then study the linear and nonlinear large-scale structure produced in our models by implementing the thin-shell estimation of the chameleon force enhancement in the spherical collapse model, which becomes mass and scale dependent in this case.
Applying the resulting effective collapse density to the Sheth-Tormen prescription, we determine the halo mass function and linear halo bias of the chameleon models.
We furthermore provide simple descriptions of the radial scalar field profile within virialised clusters using the NFW fitting function.
Based on the chameleon spherical collapse model, we introduce a mass and environment dependent chameleon modification to a mass-concentration scaling relation that is calibrated to $\Lambda$CDM $N$-body simulations.
This allows us to determine the NFW fitting parameters given the virial mass of the halo and its environmental density.
Finally, we use the halo model to describe the nonlinear matter power spectrum using our scalar-tensor modification of the linear matter power spectrum, the Sheth-Tormen halo mass function and linear halo bias, as well as the halo concentration entering the NFW halo profile.
We compare the halo model prediction against the nonlinear matter power spectrum extracted from $N$-body simulations of $f(R)$ gravity; while it provides a good qualitative description of the shape of the enhancement at high $k$-modes, it fails to recover the correct amplitude.
Introducing an effective linear power spectrum in the computation of the two-halo term that interpolates between the linear power spectrum of the chameleon model and $\Lambda$CDM and accounts for an underestimation of the chameleon suppression at intermediate scales through the linear approach, we can accurately reproduce the measurements from the $N$-body simulations over a wide range of scales.

Overall, the modelling procedures for the cosmological observables described in this paper provide useful tools to efficiently extrapolate and interpolate the nonlinear quantities extracted from $N$-body simulations beyond the simulated values of the cosmological and chameleon model parameters implemented.
Approaches of this kind are essential for the consistent study of model constraints from the observed large-scale structure, enabling sufficient and smooth variation of chain parameters as well as statistical convergence. 


\section*{Acknowledgements}

We thank Matteo Cataneo, Bridget Falck, Wayne Hu, Tsz Yan Lam, Patrick Valageas, and Gong-Bo Zhao for useful discussions.
LL and KK were supported by the European Research Council.
LL further acknowledges support from the STFC Consolidated Grant for Astronomy and Astrophysics at the University of Edinburgh and 
KK is supported by STFC (grant nos. ST/K00090/1 and ST/L005573/1) and the Leverhulme trust.
BL is supported by the Royal Astronomical Society and Durham University.
$N$-body simulations and postprocessing have been conducted on
the ICC Cosmology Machine, embedded in the DiRAC supercomputing facility funded by STFC, the Large Facilities Capital Fund of BIS, and Durham University, as well as on the Sciama High Performance Compute cluster, which is supported by the ICG, SEPnet, and the University of Portsmouth.
Further numerical computations have been performed with ${\rm Maple}^{\rm \tiny TM}~16$ and Wolfram $Mathematica^{\rm \tiny \textregistered}~9$.
Please contact the authors for access to research materials.


\vfill
\bibliographystyle{JHEP}
\bibliography{chamhalo}

\end{document}